# MAGNETIC MODELING OF INFLATED LOW-MASS STARS USING INTERIOR FIELDS NO LARGER THAN ~10 KILOGAUSS

James MacDonald and D. J. Mullan
Dept. Physics & Astronomy, University of Delaware, Newark, DE 19716, USA

ABSTRACT

We have previously reported on models of low-mass stars in which the presence of inflated radii is ascribed to magnetic fields which impede the onset of convection (e.g. MacDonald & Mullan [2017a] and citations therein). Some of our magneto-convection models have been criticized because, when they were first reported by Mullan & MacDonald (2001), the deep interior fields were found to be very large (50-100 MG). Such large fields are now known to be untenable. For example, Browning et al. (2016) used stability arguments to suggest that interior fields in low-mass stars cannot be larger than ~1 MG. Moreover, 3D models of turbulent stellar dynamos suggest that fields generated in low-mass interiors may be not much stronger than 10-20 kG (Browning 2008). In the present paper, we present magneto-convective models of inflated low-mass stars in which the interior fields are not permitted to be stronger than 10 kG. These models are used to fit empirical data for 15 low-mass stars for which precise masses and radii have been measured. We show that our 10 kG magneto-convective models can replicate the empirical radii and effective temperatures for 14 of the stars. In the case of the remaining star (in the Praesepe cluster), two different solutions have been reported in the literature. We find that one of these solutions (by Gillen et al. 2017) can be fitted well with our model using the nominal age of Praesepe (800 Myr). However, the second solution (by Kraus et al. 2017) cannot be fitted unless the star's age is assumed to be much younger (~ 150 Myr).

## 1. INTRODUCTION

In the work of Leggett et al. (2000), radii of low-mass stars were reported with sufficiently high precision that a noteworthy conclusion emerged. Although many low-mass stars were found to have radii that are consistent with the predictions of standard (non-magnetic) stellar codes, nevertheless, certain low-mass stars were found to have radii that departed significantly from the predictions. In addition, the sense of the departures was clear: the empirical radii were found to be *larger* than the predicted values. No star in the sample of Leggett et al. (2000) was found to have an empirical radius that was *smaller* than the predicted values. Subsequent studies have confirmed that some low-mass stars, but certainly not all, have radii which are inflated (beyond the error bars) relative to theoretical predictions, while there are no stars with radii smaller than predicted (beyond the error bars) (e.g. Ribas 2006).

Inspection of the data presented by Leggett et al. (2000) led Mullan & MacDonald (2001, hereafter MM01) to conclude that there is a distinguishing feature which separates stars with inflated radii from stars with predicted radii: inflated radii are more common among magnetically active stars. This suggested that radius inflation might be associated with the presence of magnetic fields in the star. In an attempt to replicate the observed inflations, MM01 obtained models of low-mass stars in which magnetic fields caused the onset of convection to be impeded. The criterion which we used to quantify the magnetic impedance of convective onset was written in terms of a single "magnetic inhibition parameter" $\delta$: this parameter had originally been derived by Gough & Tayler (1966: hereafter GT) on the basis of an energy principle which is widely used in plasma physics. The numerical value of $\delta$ is a local parameter that, at any radial location in the star, is essentially the ratio of the pressure of the vertical magnetic field to gas pressure, i.e. $\delta \approx B_v^2 / (4\pi\gamma p_{gas})$.

The principal problem that arose in obtaining the MM01 models had to do with how best to address the following issue: how does the value of $\delta$ vary as a function of radial location inside a star? For lack of any definite guidance, MM01 chose three radial profiles for $\delta$: two had $\delta$ constant at all radii, while the third had $\delta \propto [m(r)]^{2/3}$. The first two profiles lead to monotonic increases in field strength towards the center of the star. The third profile has a peak in field strength at intermediate radial locations. But in all three cases, the interior field strengths in some models are found to have values as large as tens of MG or more. With these choices of radial profiles for $\delta$, MM01 found that the most extreme field strengths emerged when we attempted to address a specific issue, namely, might it be possible for magnetic effects to suppress convection altogether in the core of a low-mass star?

1.1. *The state of knowledge in 2000-2001*

In the context of the present paper, it is pertinent to remark that the idea of complete suppression of convection in low mass stars (as discussed in MM01) presented itself in the context of certain empirical data that were available in the literature when the MM01 paper was being written (2000-2001).

The empirical data that led us to consider the above question had to do with how the coronal heating efficiency $L_X/L_{bol}$ varies as a function of spectral type. The empirical results of Fleming et al. (1993) for X-rays from M dwarfs indicated that there is no significant variation in $L_X/L_{bol}$ with spectral types between M0 and M7 (see Section 3.2 of MM01). In particular, there is no detectable "signature" in the $L_X/L_{bol}$ ratio at spectral types between M3 and M4, where the transition to complete convection (TTCC) is predicted on theoretical grounds (Limber 1958) to occur in main-sequence stars. According to Limber (1958), an interface between convective envelope and radiative core is expected to be present in the earliest M dwarfs, but is expected to be absent in the later M dwarfs. Since the interface is believed to play a significant role in dynamo activity which generates magnetic fields in a star such as the Sun (e.g. Choudhuri & Gilman 1987), it might be expected that dynamo operation could undergo a change in mode at M3-M4. A primary purpose of Fleming et al. (1993) had been to determine if the coronal X-rays contain any signature of such a change in dynamo mode. As it turned out, Fleming et al. (1993) found no evidence of any significant change in the value of $L_X/L_{bol}$ at M3-M4.

It was this empirical finding that led MM01 to consider the hypothesis that the TTCC might be postponed to later spectral types than Limber (1958) had predicted. If such a postponement actually occurs in M dwarfs, then a radiative-convective interface should persist even at spectral types later than M4. MM01 hypothesized that such an effect might occur if the magnetic field strengths were large enough to suppress convection in the core. It was our desire to generate a radiative core in M dwarfs all the way to the lower end of the main sequence that led MM01 to obtain the most extreme field strengths: we found that in order to generate radiative cores in stars with masses as small as 0.35-0.1 $M_\odot$, the models required fields of 63-110 MG.

*1.2. Theoretical constraints: equipartition?*

Browning et al. (2016) reported on a stability analysis of magnetic fields in the interior of a convective star. In the presence of effects due to buoyancy and ohmic dissipation, they found that the fields in M dwarfs can be no stronger than ~1 MG. In view of these results, the MM01 claim for the existence of fields of 63 - 110 MG in M dwarfs can no longer be regarded as physically tenable.

In fact, for some years before the work of Browning et al. appeared, we had already been obtaining successful magneto-convective models of low-mass stars in which the fields were forced to be no larger than a certain "ceiling" value: in the case of CM Dra, the ceiling field was set at a value of 1 MG (e.g. MacDonald & Mullan 2012, hereafter MM12). *A posteriori*, we consider it encouraging that the MM12 choice of an upper limit of 1MG for the fields inside M dwarfs is consistent with the subsequent conclusions of Browning et al. (2016).

In MM12, we argued that imposing an upper limit on field strength in a star is plausible because a stellar dynamo cannot be expected to generate fields that are arbitrarily strong. Might we be able to estimate an upper limit on the field strength that a dynamo can generate? A sampling of the literature in which particular dynamo models are used to predict the field strengths suggests that a definitive answer to this question may be elusive. E.g. Durney et al. (1993) model a turbulent dynamo in which a forcing term injects kinetic energy (KE) into the system and the induction equation ensures that magnetic energy (ME) is generated: the results indicate that on small scales, ME increases as time progresses, with the magnetic field approaching equipartition values $B_{eq}$ (where ME would be as large as KE).

In order to generate large-scale fields from small-scale fluctuations v and B in velocity and field, "mean-field dynamo models" rely on the "$\alpha$-effect": when averaged over large volumes, the electromotive force associated with the fluctuations, $\langle \mathbf{v} \times \mathbf{B} \rangle$, has a non-zero value which is (at small $B$) proportional to the mean B. In spherical geometry, the "$\alpha$-effect" is the source of poloidal field. However, as $B$ increases, the $\alpha$-effect is quenched (Cattaneo & Hughes 1996) to such an extent that mean fields cannot be generated which are as large as the equipartition energy.

To avoid the $\alpha$-quenching process, interface dynamo models have been developed which impose a spatial segregation between the $\alpha$-effect and the tachocline, i.e. the region of strong shear where toroidal fields are generated (e.g. Charbonneau & MacGregor 1997). In such models, $B$ values in the shear layer can exceed the equipartition field strength ($B_{eq} = 10^4$ G) in the convection zone. As an example, in a 3-D interface dynamo model, Zhang et al. (2003) find that toroidal fields can be generated in the tachocline with strengths that are 10 times larger than $B_{eq}$.

Comparison of the solar dynamo with the Earth's dynamo can highlight the widely different parameter regimes that are at work in different cosmic settings (Zhang & Schubert 2006). As an extreme example of non-equipartition dynamo activity, the Earth's magnetic field has ME/KE which has been cited as having values as large as $10^4$ (Schaeffer et al. 2017) or even $10^6$ (Roberts & King 2013).

In summary, dynamos appear to be capable of generating magnetic fields that are either weaker or stronger than $B_{eq}$, depending on particular conditions. In Section 2 below, we refer to 3-D models of stellar convection zones where particular values of maximum $B$ (= 13 - 14 kG) have been obtained. In view of the discussion in the present section, we will not refer to equipartition arguments in evaluating these $B$ values: we will simply adopt their numerical values as a practical limit on field strengths inside a star.

1.3. *Empirical constraints in 2017*

From an entirely different perspective, recently published empirical results (Houdebine et al. 2017) have indicated that the use of *coronal* signatures is not conducive to detecting the effects of the TTCC among M dwarfs. Competing effects conspire to obscure any coronal signature. It is preferable to use *chromospheric* signatures, where a change in slope of the rotation-activity correlation (RAC) has been identified between dM2e and dM3e stars. Houdebine et al. (2017) argue that this change in slope may be a signature of the TTCC.

In view of these results, we now recognize that MM01 did not need to interpret the coronal results of Fleming et al. (1993) to indicate that convection was suppressed in the cores of stars of spectral type M3-M4 and later. From this point of view, the MM01 claim that fields of 63-110 MG are present in stars with masses of 0.35-0.1 M☉ was unnecessary.

One further constraint has emerged in the recent work of Kochukhov & Lavail (2017) who report a "remarkable difference" in the global field topology in two M dwarfs (UV Cet, BL Cet) in a binary with essentially identical spectral types (M5.5-6) and rotational periods (0.23-0.24 days). Using circular polarization data, one component is found to have a mean line-of-sight field that is 4 times stronger than the other, and the stronger field is dipolar while the weaker field is non-axisymmetric. But, when Zeeman broadening is used to estimate the total magnetic flux, both components are found to have comparable <$B$>, namely 5-6 kG. Despite the "surface" differences in field topology, the two stars are observed to have similar radii, suggesting that the *interior* field strengths in both components are comparable. Unfortunately, our modeling approach is one-dimensional, and so it does not allow us to determine why the geometry of the field might be different in two otherwise similar stars.

1.4. *Goal of the present paper*

Unfortunately, the strong fields reported in MM01 remain in the literature as a reason for readers to suggest that our magneto-convective models are unrealistic in the context of low-mass stars. E.g. Kraus et al. claim: "Magnetism only seems to change the stellar parameters of fully convective main sequence stars if interior magnetic field strengths are far higher than expected (B >1 MG)."

In the present paper, our goal is to challenge this criticism, and show that it is unfounded. Specifically, we wish to expunge the negative connotations associated with the untenably large fields that we reported in MM01. We wish to demonstrate that the observed properties of inflated low-mass stars (radii, mass, temperature, luminosity) *can* be replicated by our magneto-convective models in which the interior magnetic field strengths are certainly consistent with the 1 MG upper limits of Browning et al. (2016). But in order to strengthen our case even more, we set upper limits on interior field strength in the present paper which are 100 times smaller than the Browning et al. (2016) limits.

Although there are some indications that radius inflation occurs in old and inactive single stars (Spada et al. 2013; Mann et al. 2015), here we focus on stars in binary systems for which masses have been measured with an accuracy that, in general, far exceeds that which can be obtained for single stars. Furthermore, the systems that we consider here contain only M dwarfs, which are either fully convective or have deep surface convections. Consequently, we did not need to consider complications that might arise, for example, from heavy element settling in G stars, which have shallower surface convection zones. In regard to whether inactive stars show radius inflation, we note that there are a number of binary systems that contain inactive stars and their radii are consistent with models that do not include magnetic effects. Examples are KOI 126 (Carter et al. 2011; Feiden, Chaboyer & Dotter 2011), KIC 6131659 (Bass et al. 2012) and T-Cyg1-12664 (Han et al. 2017). These objects support the idea that radius inflation is associated with activity and by inference the presence of magnetic fields, and is not due to uncertainties associated with, for example, choice of equation of state, unaccounted opacity sources, helium abundance, and heavy element abundances (see MM12 for a detailed analysis of how such uncertainties affect radius inflation in the context of CM Dra).

Surface dark spots are another manifestation of magnetic fields and are found or inferred to be present on a number of active stars [e.g. the components of YY Gem (Torres & Ribas 2002) and CM Dra (Morales et al., 2010)]. It has been shown that the presence of surface dark spots increases stellar radii (Spruit 1992; Chabrier et al. 2007). On the other hand, polar spots complicate eclipse modeling and may lead to overestimation of stellar radii (Morales et al. 2010; Kraus et al 2011). Both spot effects were considered in the case of CM Dra by MM12 and MM14, who found that both spots and magnetic inhibition of convection contributed to the observed radius inflation. MM12 showed that, for the theoretical models of fully convective stars, there is degeneracy in radius inflation from the effects of reducing the mixing length ratio and radius inflation from inclusion of dark spots. In addition, MM14 showed that there is a similar degeneracy between the effects of reducing the mixing length ratio and increasing the degree of magnetic inhibition of convection provided interior field do not exceed $10^7$ G. Because of the difficulty of disentangling the effects of dark spots, choice of mixing length ratio, and magnetic inhibition of convection, we have chosen to ignore the effects of dark spots in the analysis of the three new binary systems that we consider here.

### 1.5. *Comparison and contrast with the model of Feiden and Chaboyer*

Feiden & Chaboyer (2012, hereafter FC; and 2013, hereafter FC13) have presented an independent approach to modeling the effects of magnetic fields on convection in stars. The FC approach overlaps in some respects with our approach. For example, FC concentrate on the radial component of the magnetic field, since that is "the component necessary for stellar evolutionary computations" (their Section 3.1). In our application of the GT criterion, the relevant magnetic field is the vertical component, which in 3-D is equivalent to the radial component. Also FC use an equation for magneto-hydrostatic equilibrium (their eq. 13) which uses the gradient of magnetic plus gas pressure, as MM do also (Section 4.5.2 in MM01). But FC also include effects which differ from our treatment. We describe these differences in the sub-sections 1.5.1-1.5.5.

### 1.5.1. *Criterion for onset of convection*

A major difference between FC and MM is the inclusion (by FC) of a thermodynamic state variable $\chi$, which is defined as the magnetic energy per unit mass. This variable alters the equation of state in such a way that a fluctuation in density $d\rho/\rho$ is related not only to $dp/p$ and $dT/T$, but also to $d\chi/\chi$: FC find that $d\rho/\rho$ includes a term that has the form $-v\, d\chi/\chi$ where the coefficient $v$ is determined by how sensitive $\rho$ is to changes in $\chi$. As it turns out, the coefficient $v$ is essentially equal to the MM magnetic inhibition parameter $\delta$ (see FC13). In the presence of a vertical magnetic field, MM require that, in order for convection to set in, the temperature gradient must exceed the adiabatic gradient by a finite amount, namely, $\delta$. (This result was derived by GT in a medium of uniform density.) In contrast, FC find that in order for convection to set in inside a star (a medium where density is not uniform), the excess of the temperature gradient above adiabatic does not need to be as large as $\delta$ ($\approx v$). Instead, the excess needs to equal $\delta$ ($\approx v$) times the logarithmic gradient $g_\chi = d\log\chi/d\log p$. Numerically, FC find that $g_\chi$ is of order 0.1 (see FC13). Thus, a given magnetic field does not inhibit convection as seriously in the FC model as in the MM model. As a result, in order for magnetic fields to alter stellar structure to a particular extent (i.e. in order to inflate the radius of a particular star), the FC model is required to use stronger fields than MM require. Quantitatively, FC13 state: "we expect our required magnetic field strengths to be about a factor of 2-5 larger" than those required by MM.

In view of this, comparison between model results and empirical measures of field strengths on the surfaces of low-mass stars might help to decide which models are better. However, this is not a clear-cut test, because FC calculate their models by treating the surface field strength $B_{surf}$ as a boundary condition, i.e. a free parameter. FC state that $B_{surf}$ "has the potential to be constrained observationally". In contrast, MM make no assumption about $B_{surf}$ as a boundary condition: instead, MM use the output from a model to *derive* a value for $B_{surf}$ in each star. In view of this, observations of $B_{surf}$ provide a significant *a posteriori* test of the MM models. So far, we have found that MM predictions of $B_{surf}$ are consistent with a variety of empirical constraints (MacDonald & Mullan 2014, hereafter MM14; MacDonald & Mullan 2017b, hereafter MM17b). Interestingly, the strongest value we have obtained for $B_{surf}$ is 1.4 kG (MM17b), whereas FC cite an example of a surface field of 5 kG. This difference in field strengths falls within the range of 2-5 mentioned by FC13.

1.5.2. *The choice of radial profile for the magnetic field*

The second major difference between MM and FC has to do with the (unknown) distribution of field strength inside a star. In the earliest models, MM01 choose to explore two models with $\delta$ = constant at all radii, and one model with $\delta$ declining monotonically from surface to center. More recently (MM12), $\delta$ has been assumed to remain constant in the outer layers of the star, but then, starting at a certain radial location $r_c$, the field strength is capped at a fixed value (the "ceiling") at all (interior) $r$ values from $r_c$ to $r = 0$.

Examples of our magnetic field and $\delta$ profiles are shown in figures 1 and 2 for a solar composition 0.35 $M_\odot$ model at age 5 Gyr. Here $z$ is depth measured from the surface of the star.

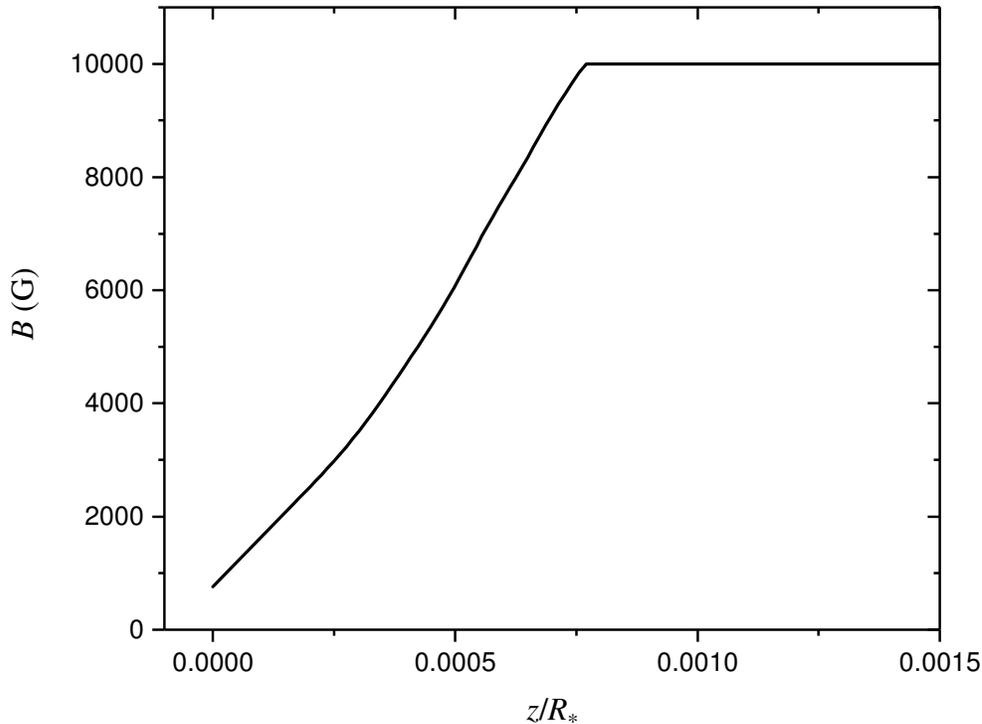

Figure 1. Representative magnetic field profile $B(r)$ in the outermost layers of a star in which the field ceiling is $10^4$ G.

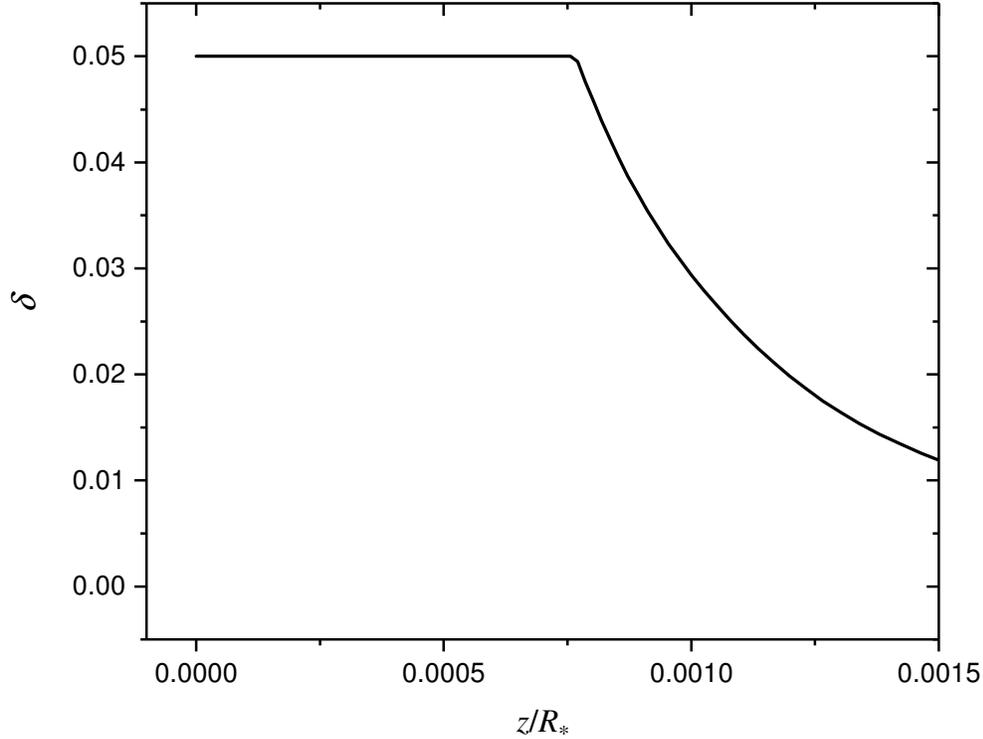

Figure 2. Profile of the magnetic inhibition parameter $\delta$ corresponding to the magnetic field profile shown in figure 1.

As regards the selection of a radial profile of the field, FC admit that "without any real confidence of the radial profile inside stars, we are left to our own devices". They select a profile which they refer to as a "dipole configuration" such that the field has maximum strength at some radial location, $r = r_d$. This location corresponds either to the base of the convective envelope, or to a radial location where maximum field strength is generated by a turbulent dynamo. Once $r_d$ is chosen, FC assume that $B(r)$ falls off as $r^{-3}$ at $r > r_d$, while the interior $B(r)$ increases as $r^3$ in the range $0 < r < r_d$.

Although FC do not say this explicitly, it is possible that their choice of $r^{-3}$ variation for the exterior field (i.e. $r > r_d$) may be associated with the well-known results that outside a uniformly magnetized sphere, or far from a current loop, the field strength decreases as $r^{-3}$. However, contrary to the $r^3$ dependence of $B(r)$ assumed by FC as $r \rightarrow 0$, the field inside a magnetized sphere, or inside a current loop, does not go to zero at the center: on the contrary, the interior field has a constant strength (e.g. Jackson 1962).

1.5.3. *Radial profile of magnetic field in the MM model: the deep interior*

In the MM model, when we consider conditions starting at the surface of the star, the field is at first allowed to increase in strength as we consider regions that lie deeper below the surface (see Fig. 1). The increase in field strength in this outer region is assumed to be such that the numerical value of $\delta \sim B^2/p$ retains its boundary value. To achieve this, $B$ must increase in proportion to $\sqrt{p}$ as we go deeper below the surface. However, this increase in *B* at increasing depth does not continue indefinitely inside the star. Instead, the increase in *B* stops at a certain radial location $r = r_c$ where the field reaches its "ceiling" value $B_c$. At all radial locations from $r = r_c$ inwards to the center of the sphere, our models assume that *B* remains constant at the value $B_c$ (see Fig. 1). Our choice of radial profile for *B(r)* interior to $r = r_c$ replicates the physically expected behavior of field strength in the interior of a uniformly magnetized sphere (Jackson 1962): $B(r) = B_c$ remains constant throughout the interior, $0 < r < r_c$.

Our deep interior radial profile of *B(r)* is quite different from that of FC. For FC, *B(r)* does not retain a constant value throughout the interior of the star, but instead, *B(r)* falls off rapidly to zero as $r \rightarrow 0$.

1.5.4. *Radial profile of magnetic field in the MM model: the surface layers*

Let us see if we can determine the functional form of the radial profile for *B* in the outer layers of an MM model, i.e. between $r = r_c$ and the surface $r = R$. In these regions, convection is at work, and so *dT/dz* is of order $g/c_p$, where $c_p$ is the specific heat at constant pressure, and *z* is depth measured from the stellar surface. In regions where the gas is monoatomic and non-ionizing, *dT/dz* = constant, and so *T* ~ *z*. To determine $B(z) \sim \sqrt{p(z)}$ in layers where $\delta$ = constant, we need to know how *T* is related to *p*. If the adiabatic gradient $\nabla_{ad} = 2/5$, as for an ideal gas, adiabatic processes have $p \sim T^{5/2}$, and *B(z)* would increase as $T^{5/4}$, i.e. $B \sim z^{5/4}$. Therefore, as we go deeper into a model, *B* will increase in strength, but only slightly faster than a linear function of *z*.

The essence of a convection zone is that gas is undergoing ionization. In such conditions, $\nabla_{ad}$ is no longer as large as 2/5: in the case of pure hydrogen, $\nabla_{ad}$ can be as small as 0.1 (e.g. MacDonald 2015). In the case of the solar convection zone, where H and He are present, $\nabla_{ad}$ falls to a value of 0.16 (e.g. Mullan 2009). In such conditions, *p* becomes a much steeper function of *T*: $p \sim T^{6.3}$. Therefore $B(z) \sim z^{3.15}$, i.e. $d \log B / d \log z = 3.15$. This is the steepest variation of *B(z)* that we expect to find in the convection zone. Comparing with the slope of 5/4 for non-ionizing gas, we expect that in the regions of our models where $\delta$ = constant, the field should vary with depth according to a (logarithmic) slope in the range 1.25 - 3.15.

1.5.5. *How do the FC models and ours compare?*

There are some overlaps between the MM and FC models, and some differences. Successful fits of models to a set of observational data, including $B_{surf}$ values, may help to decide which model (if any) is more realistic.

## 2. A TWO-PARAMETER FAMILY OF MAGNETO-CONVECTIVE MODELS

In several papers which we have published since MM01 appeared, various choices for the radial profile of $\delta$ have been explored which are very different from those used in MM01. In particular, as mentioned in Section 1.2 above, we have chosen radial profiles of a kind that have precisely the following property: the magnetic field in the stellar interior is *not allowed to exceed a value* that we consider appropriate for low-mass stars. We refer to this upper limit on the interior field as the "ceiling" field $B_{ceil}$. In contrast to the one-parameter ($\delta$) models, which were presented by MM01, we have, in more recent papers, been developing families of models based on the two-parameters: $\delta$ and $B_{ceil}$.

To reduce the size of parameter space, it would be constructive if we could identify physical reasons that would allow us to set limits on the value of the parameter $B_{ceil}$. Browning et al (2016) have pointed out an upper limit, of order 1 MG. (We have already explored parameter space in the vicinity of that upper limit: MM12, MM14, MM17a, b). Here we would like to consider a lower limit on $B_{ceil}$. We note that 3D numerical dynamo models obtained by Browning (2008) find that the fields which can be generated by turbulent convective motions in low-mass stars have a maximum field strength of 13.1 kG in a model which rotates with the same period as the Sun (~27 days). More recently, Yadav et al. (2015) have reported that the maximum field strength in a completely convective star with a rotation period of 20 days is 14 kG. To be sure, M dwarfs (especially the active stars) are known to have rotation periods that are shorter than 20 days by factors of at least 2 (see Houdebine & Mullan 2015: their figure 10). In the case of such faster rotators, fields stronger than 14 kG might be expected to be generated (see, e.g. MM12). In view of these results, it seems that values of 10 -20 kG could serve as plausible values for $B_{ceil}$.

To be conservative, in the present paper, we explore the following question: if we were to adopt $B_{ceil}$ = 10 kG, could we obtain successful models for the inflated radii which have been reported for a sample of more than a dozen low-mass stars in the literature?

We claim that our choice of $B_{ceil}$ = 10 kG in the present paper should be regarded as a plausible *lower* limit on the strength of the peak fields *inside* active M dwarfs. The reason for this claim is that surface fields in M dwarfs with spectral types of M4.5e and earlier have already been reported to be 1-4 kG (Saar 1996). At later spectral types, even stronger surface fields have been found. E.g. In a study of M5-M6.5 stars, Shulyak et al. (2017) report surface fields as large as 7 kG. In view of these results, it seems to us that it would be unphysical to consider peak fields in the *interior* of an active star that would be *smaller* than the field strengths which are known to exist on the surface.

## 3. PREVIOUS RESULTS USING TWO-PARAMETER SOLUTIONS

### 3.1. *CM Dra*

The second parameter $B_{ceil}$ was introduced for the first time (in MM12) into our models in the context of CM Dra. We argued that in CM Dra, where the rotational angular velocity $\Omega$ is 22 times faster than solar, dynamo activity *might* generate fields as strong as 1 MG: accordingly, we used $B_{ceil}$ = 1 MG to fit CM Dra in MM12. This choice of $B_{ceil}$ is not unphysical (Browning et al. 2016). But in the present paper, in case dynamos cannot generate 1 MG fields, we revisit CM Dra to see if the data are also consistent with smaller $B_{ceil}$.

### 3.2. *Surface magnetic field strengths: lack of sensitivity to $B_{ceil}$*

Our magnetic models were successfully used to replicate estimates of the vertical component of the magnetic field $B^V_{surf}$ on the surfaces of three low-mass stars with parameters that are among the most precisely known among low-mass stars (MM14). In MM14, we explicitly illustrated the two-parameter nature of our solutions by deriving acceptable solution paths through the 2-parameter space ($B_{ceil}$, $\delta$). An interesting aspect of our solutions emerged: even though $B_{ceil}$ was allowed to vary by a factor of $10^4$, the solutions for $\delta$ were found to vary by no more than ~2. Thus, our solutions for $B^V_{surf}$ (which are ~ $\sqrt{\delta}$) are very insensitive to our choice of $B_{ceil}$.

### 3.3. *Quantitative dependence of surface field strength on $B_{ceil}$*

MacDonald & Mullan (2017a) quantified this insensitivity: it was found that our solutions were fitted by $B^V_{surf} \sim B_{ceil}^{-0.07}$. Thus, if we choose to *reduce* $B_{ceil}$ by a factor of (say) 100, then the value of $B^V_{surf}$ must *increase* by 1.4 in order to replicate a given amount of radius inflation.

Thus, even if $B_{ceil}$ (an unobservable quantity) is permitted to take on any value in a broad range, the resulting solution has a narrowly constrained value of $B^V_{surf}$ (an observable quantity). We consider it an important feature that the best constrained aspect of our solution ($B^V_{surf}$) is in principle testable.

### 3.4. *Testing our solutions for $B^V_{surf}$ using radio emission*

LSPM J1314 is the strongest radio emitter among ultra-cool dwarfs. The radio emission could be either gyrosynchrotron emission (GSE) or electron cyclotron maser (ECM) (McLean et al. 2011). In MM17b, magnetic models of both components were obtained. These models used values of $B_{ceil}$ ranging from $10^4$ G to $10^6$ G, leading to solutions with values of $B^V_{surf}$ ranging from 630 to 1430 G. These $B^V_{surf}$ values are large enough that they suggest ECM as the preferred mode of radio emission.

### 4. MAGNETIC MODELS OF SIX INFLATED LOW-MASS STARS IN BINARIES

We now turn to a demonstration of how, without relying on "extreme" fields, we can obtain successful magnetic-convective models of 15 stars using $B_{ceil}$ = 10 kG. We start (in Sections 4.1-4.3) with 6 stars in three eclipsing binaries which were reported recently by independent groups. Then in Section 4.4, we re-visit 9 stars for which we obtained (in earlier papers) magnetic models with large $B_{ceil}$ and ask: can we obtain satisfactory fits to the same empirical radii and effective temperatures using $B_{ceil}$ = 10 kG? Our modelling technique has been recently described in MM17b. We begin by constructing nonmagnetic models of appropriate mass and age that use BT-Settl atmosphere models to provide the outer boundary conditions. Because the BT-Settl atmospheres do not include magnetic effects, we next construct non-magnetic models using Krishna-Swamy (1966) atmospheres that match the models which use the BT-Settl atmosphere boundary conditions, adjusting the mixing length parameter when necessary. Finally, we construct models that include magnetic effects consistently in the Krishna-Swamy atmosphere and the stellar interior. In all models described here, we assume that magnetic diffusivity effects can be neglected, i.e. we assume that during the (short) lifetime of a convective cell in a low mass star, the magnetic field drifts relative to the gas by a length which is small compared to the cell size.

4.1. *LP 661-13*

This M3.5 object was discovered to be a double-M binary in MEarth records (Dittmann et al. 2017). Masses and radii of both components were reported with precisions of 1% or better. The primary, LP 661-13A, has mass 0.30795 ± 0.00084 $M_\odot$ and radius 0.3226 ± 0.0033 $R_\odot$ while the secondary, LP661-13B, has mass 0.19400 ± 0.00034 $M_\odot$ and radius 0.2174 ± 0.0023 $R_\odot$. Dittman et al. reported that both stars appear inflated compared to standard stellar models. However, the primary star was found to be significantly more inflated compared to standard models of stars (with the appropriate mass) than the secondary star. Moreover, attempts by Dittmann et al. to fit the photometry of the system indicated that the photometric solution is best if dark spots are placed on the primary. Spots on the secondary do not give an equally satisfactory photometric solution.

To test our magnetic models on this system, we used the limit on its age of > 4.3 Gyr cited by Dittmann et al. based on circularization of the orbit. Dittmann et al. estimate that [Fe/H] = 0.0 for the primary star and Fe/H] = -0.13 for the secondary star, with an uncertainty of approximately 0.1. For our models, we adopt solar composition.

Results for standard (non-magnetic) models of primary and secondary are shown in figure 3. In the figure, the measured radii (including error bars) are indicated by dashed horizontal lines. Our non-magnetic evolutionary models are indicated by solid curves. We see that both stars have measured radii that are larger than the model predictions. However, it is also clear from figure 3 that the secondary star has a radius that almost overlaps with the standard model at large age. In the case of the primary star, the observed radius is significantly larger than the non-magnetic model predicts at any main-sequence age. The primary star, according to our interpretation, needs a magnetic field that is strong enough to alter significantly the structural properties of the star, i.e. to increase its radius by up to 8% relative to a non-magnetic model.

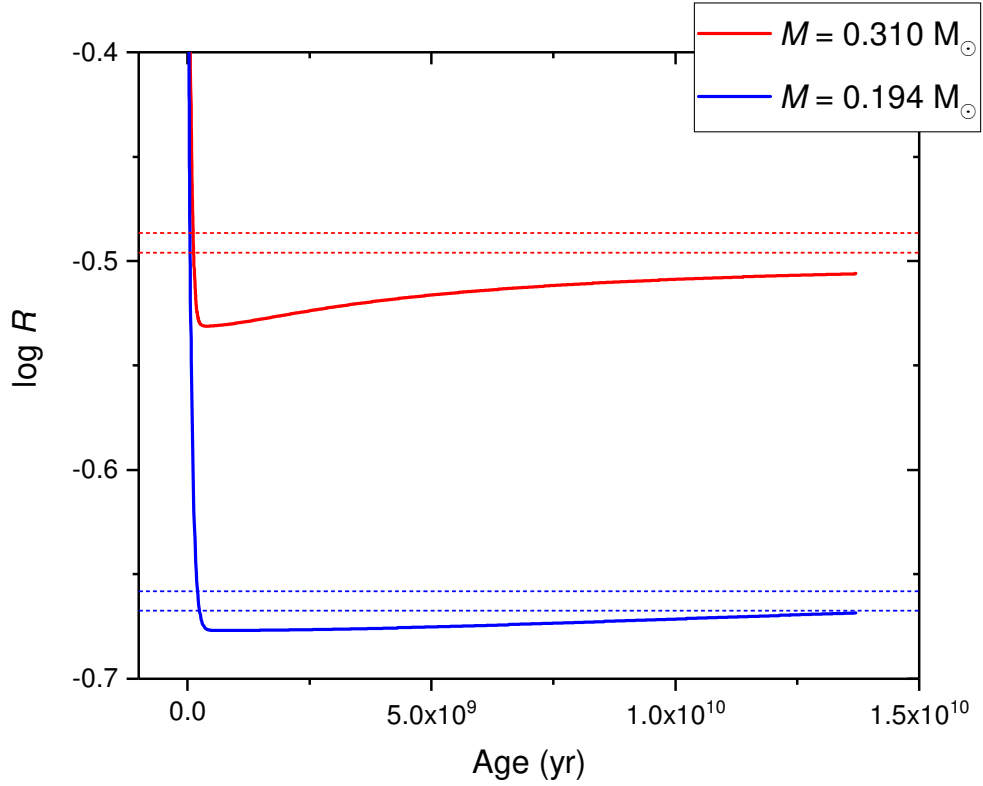

Figure 3. The solid lines show the evolution of stellar radius for non-magnetic models of the primary (red) and secondary (blue) components of LP 661-13. The broken lines show the $1\sigma$ limits on the radii determined by Dittmann et al. (2017).

In order to fit the empirical radii, we have computed magneto-convective models for each component. We have performed Monte Carlo simulations to determine the distribution of $\delta$ values that provide fits to the stellar radii. We have assumed that the age has a uniform distribution between 4.3 and 13.7 Gyr, and the measured radius has a normal distribution with the mean and standard deviation found by Dittman et al. For models using $B_{ceil}$ = 10 kG, we find $\delta = 0.095 \pm 0.013$ (in the outer layers) for the primary. The distribution of corresponding surface vertical magnetic field has $B^{V}_{surf} = 1260 \pm 80$ G. For the secondary star, we find $\delta = 0.038 \pm 0.015$ and $B^{V}_{surf} = 830 \pm 250$ G. Thus, we find that the primary component has the stronger field. This is consistent with (i) the larger inflation in radius of the primary as seen in Fig. 1, and (ii) Dittman et al.'s discovery that photometric solutions call for star spots on the primary, rather than on the secondary.

It appears that we can successfully replicate the two components of LP 661-13 with magneto-convective models of main sequence stars in which the field strength is no larger than 10 kG.

## 4.2. KELT J041621-620046

This is a double-M dwarf eclipsing binary with an orbital period of ~1.11 d (Lubin et al. 2017). The component masses and radii are $0.447^{+0.052}_{-0.047}$ and $0.399^{+0.046}_{-0.042}$ $M_\odot$, and $0.540^{+0.034}_{-0.032}$ and $0.453 \pm 0.017$ $R_\odot$. Lubin et al. report "Each star is larger by 17-28% and cooler by 4-10% than predicted by standard (non-magnetic) stellar models". Lubin et al. estimate that [M/H] = -0.2±0.2 from the $\zeta_{TiO/CaH}$ index (Reid et al. 1995) using the calibration from Lépine et al. (2013). The values of $T_{eff}$ are determined to be 3481 ± 83 and 3108 ± 75 K, for primary and secondary respectively.

Both components have strong emission in Hα, indicating chromospheric activity in both stars. Strong activity suggests that magnetic fields are present.

Lubin et al. find that orbit has a small but non-zero eccentricity, $e = 0.034 \pm 0.011$. For main sequence models of representative age 5 Gyr, we estimate by solving the equations given by Zahn (1989) that the circularization timescale is ~2 $10^8$ yr, which suggests that either the system is relatively young or there is a third body perturbing the orbit. Hence, we need to consider both main sequence and pre-main sequence scenarios.

### 4.2.1 *Main Sequence Scenario*

Because the mass estimates have relatively large statistical errors, we have computed solar composition models for masses from 0.30 to 0.55 $M_\odot$ in increments of 0.05 $M_\odot$. We show in figure 4, $R$ plotted against $T_{eff}$ at an age of 5 $10^9$ yr for $\delta$ values of 0.00, 0.08 and 0.16. The grey arrow indicates the shift in the isochrones between solar composition models and models with [Fe/H] = -0.2.

We see that agreement between observation and modelling is obtained provided the stellar masses are near their 1$\sigma$ upper limits. The estimated surface vertical field strengths from solar composition models are 700 – 810 G for the primary and 1080 – 1300 G for the secondary.

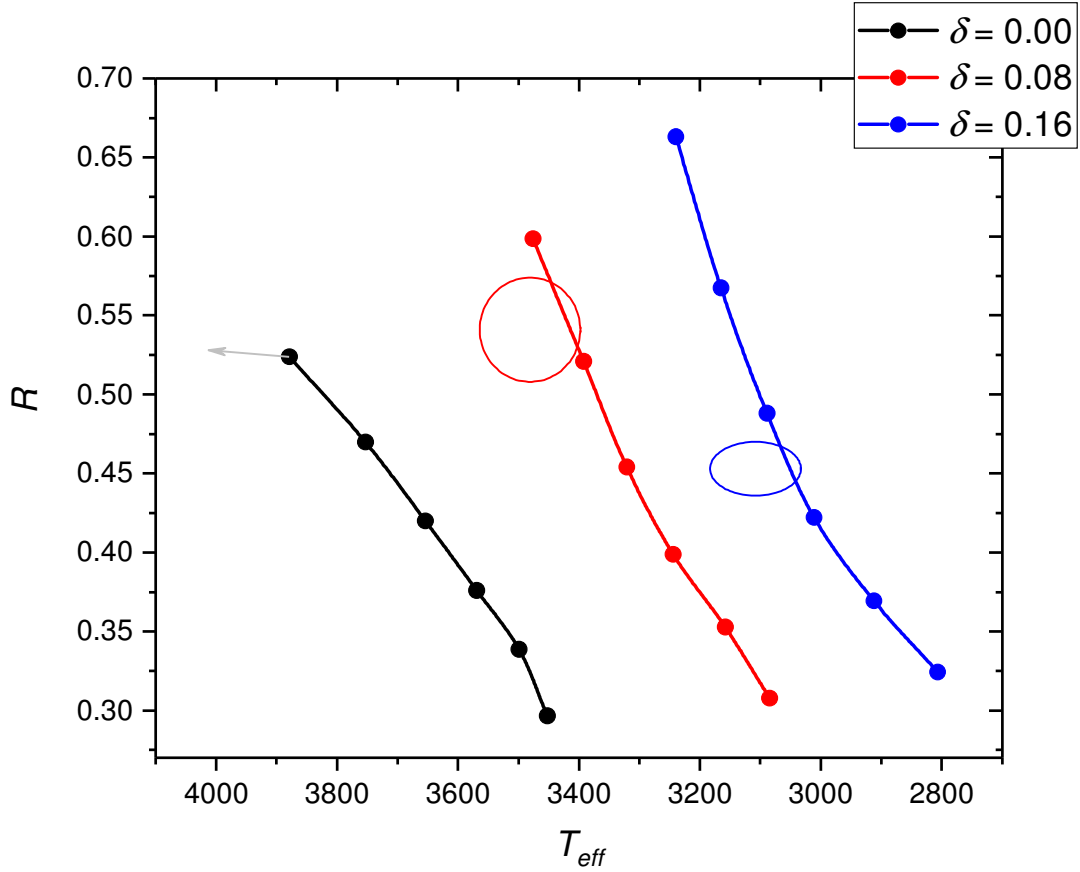

Figure 4. Location of the error ellipses for the primary (red) and secondary (blue) components of KELT J041621-620046 in the $T_{eff}$ – $R$ plane. Also shown are 5 Gyr isochrones for $\delta = 0.00$ (black), 0.08 (red) and 0.16 (blue). The filled circles correspond to mass values ranging from 0.30 to 0.55 $M_\odot$ in increments of 0.05 $M_\odot$. The grey arrow indicates how the isochrones would move if the heavy element abundance were reduced from [Fe/H] = 0.0 to Fe/H] = -0.2.

### 4.2.2 *Pre-Main Sequence Scenario*

The primary is consistent with non-magnetic models for ages 3 - 4 x $10^7$ yr. At this age, non-magnetic models predict for the secondary a radius (within the 1$\sigma$ errors) that is too large by 0.015 $R_\odot$ and $T_{eff}$ too high by ~300 K. Models consistent with the observed $R$ and $T_{eff}$ can be found for both components at age ~8 $10^7$ yr by including magnetic effects. For the primary we find $\delta = 0.038 \pm 0.015$, and for the secondary $\delta = 0.088 \pm 0.013$. The corresponding ranges in magnetic field strength are $B_{surf} = 520 \pm 110$ G, and $1080 \pm 80$ G. Figure 5 compares the locations of the two components of KELT J041621 in the $T_{eff}$ - $R$ diagram with 80 Myr isochrones for models with $\delta = 0$, 0.04 and 0.08.

It appears that we can successfully replicate the two components of KELT J041621with magneto-convective models of pre-main sequence stars in which the interior fields in both components are no larger than 10 kG.

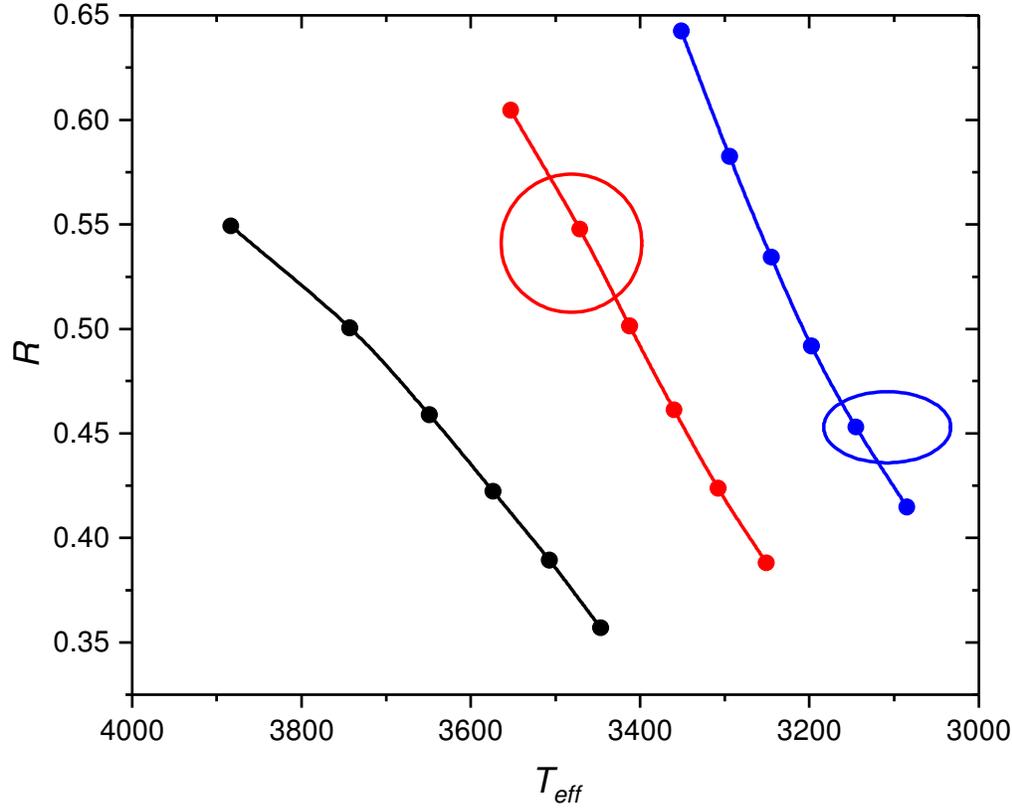

Figure 5. Location of the error ellipses for the primary (red) and secondary (blue) components of KELT J041621-620046 in the $T_{eff} - R$ plane. Also shown are 80 Myr isochrones for $\delta = 0.00$ (black), 0.04 (red) and 0.08 (blue). The filled circles correspond to mass values ranging from 0.30 to 0.55 $M_\odot$ in increments of 0.05 $M_\odot$.

### 4.3. AD 3814 (=PTFEB132.707+19.810)

This is a 6.0 d double line eclipsing binary in the Praesepe cluster that has been investigated recently by a number of groups, including Gillen et al. (2017) and Kraus et al. (2017). Gillen et al. determine that primary has mass, $M_1 = 0.3813 \pm 0.0074$ $M_\odot$ and radius $R_1 = 0.3610 \pm 0.0033$ $R_\odot$. For the secondary they find mass, $M_2 = 0.2022 \pm 0.0045$ $M_\odot$ and radius, $R_2 = 0.2256^{+0.0063}_{-0.0049}$ $R_\odot$. Kraus et al. find a similar mass and radius for the primary, $M_1 = 0.3953 \pm 0.0020$ $M_\odot$ and $R_1 = 0.363 \pm 0.008$ $R_\odot$. They also find a similar mass for the secondary, $M_2 = 0.2098 \pm 0.0014$ $M_\odot$, but their radius measurement, $R_2 = 0.272 \pm 0.012$ $R_\odot$, is significantly higher than that of Gillen et al. Both groups have determined the effective temperatures from spectroscopy. Gillen at al. find $T_{eff,1} = 3211^{+54}_{-36}$ K and $T_{eff,2} = 3103^{+53}_{-39}$ K. Kraus et al. find similar temperatures with $T_{eff,1} = 3260 \pm 30 \pm 60$ K and $T_{eff,2} = 3120 \pm 50 \pm 60$ K.

Kraus et al. state that the primary is found to be cooler and less luminous than standard models, while the secondary is cooler and substantially larger (by 20%) than standard models. However, the secondary radius found by Gillen et al. is consistent with standard models. In view of the large difference in secondary radius determination, we analyze the results of the two groups separately.

The age of the Praesepe cluster is estimated to be 600–800 Myr (Kharchenko et al. 2005; Delorme et al. 2011; Brandt & Huang 2015). A range of heavy element abundance measurements appear in the literature. Friel & Boesgaard (1992) determined from analysis of 6 F dwarfs that Praesepe has a near solar composition with [Fe/H] = 0.038 ± 0.039. This is supported by Taylor (2006) who found [Fe/H] = 0.01 ± 0.04. However Pace et al. (2008) derived from seven solar-type stars a super-solar heavy element abundance of [Fe/H] = 0.27 ± 0.10. Intermediate values of the heavy element abundances have also been found. From 11 solar-type stars, Boesgaard et al. (2013) derived [Fe/H] = 0.12 ± 0.04, and from three red giant stars Carrera & Pancino (2011) obtained [Fe/H] = +0.16 ± 0.05, which is consistent with the result of Boesgaard et al. (2013). More recently, Yang, Chen & Zhao (2015) derive [Fe/H] = 0.16 ± 0.06 from 4 giant stars, including the 3 stars analyzed by Carrera & Pancino (2011), and for other stars in the cluster find a range of [Fe/H] values ranging from 0.048 to 0.20.

In view of the large range of abundance determinations, we have made models for [Fe/H] = 0.0, 0.1 and 0.2.

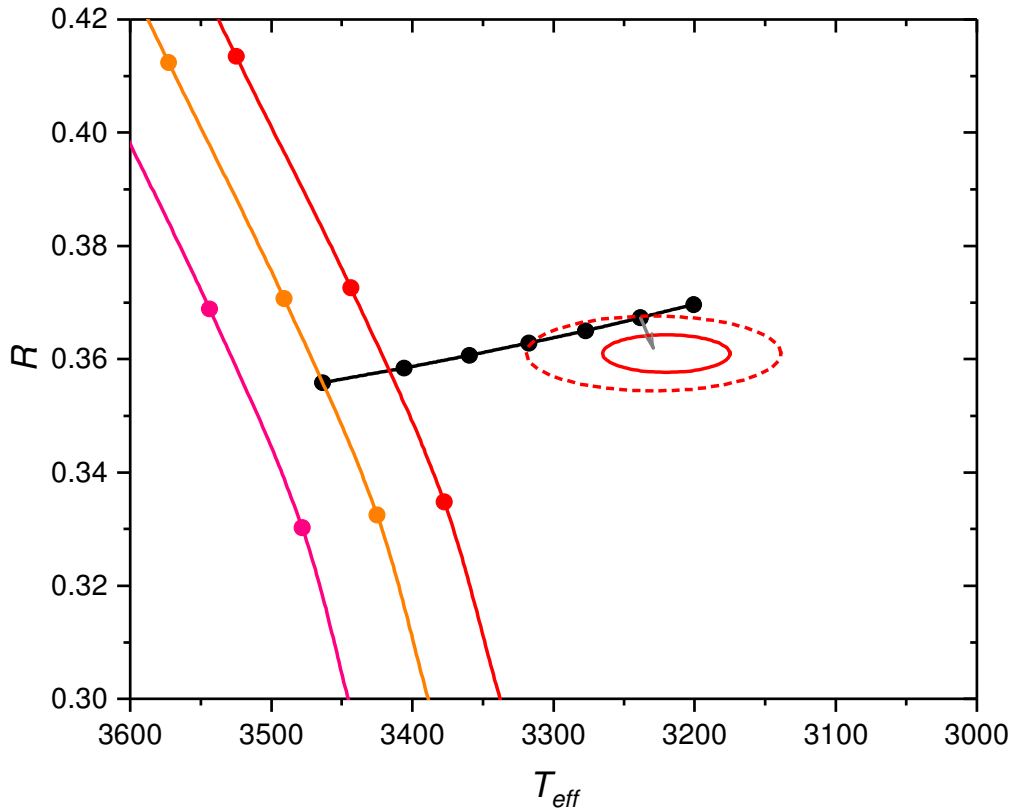

Figure 6. Location of the primary component of AD 3814 in the $T_{eff}$ - $R$ diagram, using the results of Gillen al. (2017). The solid and broken ellipses show the 1 and $2\sigma$ errors, respectively. Also shown are the non-magnetic 800 Myr isochrones for [Fe/H] = 0.0 (pink), 0.1 (orange) and 0.2 (red). The black line shows how the 0.381 $M_\odot$ [Fe/H] = 0.1 isochrone moves as $\delta$ is increased from 0.0 to 0.06 in steps of 0.01. The grey arrow shows how this isochrone would shift if the mass were equal to its $1\sigma$ lower limit of 0.374 $M_\odot$.

4.3.1 *Analysis based on Gillen et al. measurements*

In figure 6, we show the location of the primary component of AD 3814 in the $T_{eff}$ - $R$ diagram using the results of Gillen et al., together with the non-magnetic 800 Myr isochrones for [Fe/H] = 0.0, 0.1 and 0.2. The solid and broken ellipses show the 1 and $2\sigma$ errors, respectively. The black line is the 800 Myr isochrone for magneto-convective models of mass 0.381 $M_\odot$ as $\delta$ is increased from 0.0 to 0.06 in steps of 0.01. Also shown is how this isochrone would move if the mass were reduced to its $1\sigma$ lower limit of 0.374 $M_\odot$.

We see that the magneto-convective mean mass [Fe/H] = 0.1 models of the primary component match the data at the $2\sigma$ level for $\delta$ values between 0.030 and 0.050. The corresponding range for the vertical component of the magnetic field strength at the surface of the primary component is 490 - 670 G. If the mass is equal to the $1\sigma$ lower limit, then the models match the data at the $1\sigma$ level for $\delta$ values between 0.041 and 0.058. The corresponding range in surface field strength is 580 – 720 G.

Figure 7 is the same as figure 6 but for the secondary component of AD3814.

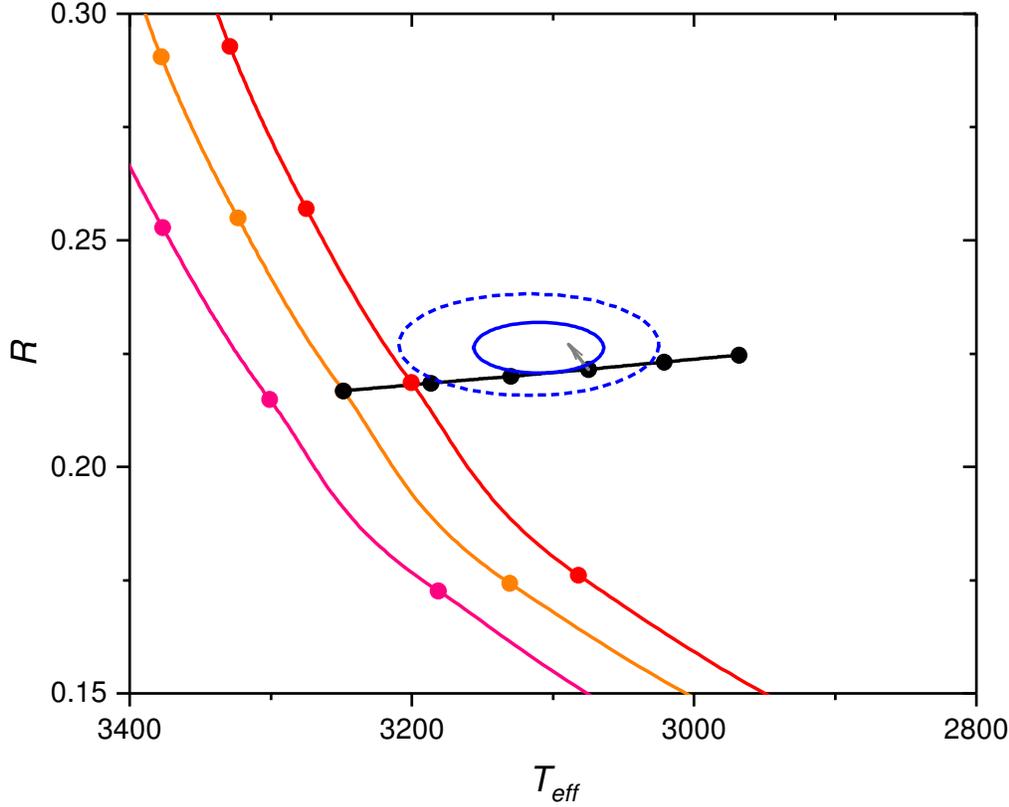

Figure 7. Location of the secondary component of AD 3814 in the $T_{eff}$ - $R$ diagram, using the results of Gillen et al. (2017). The solid and broken ellipses show the 1 and $2\sigma$ errors, respectively. Also shown are the non-magnetic 800 Myr isochrones for [Fe/H] = 0.0 (pink), 0.1 (orange) and 0.2 (red). The black line shows how the 0.200 $M_\odot$ [Fe/H] = 0.1 isochrone moves as $\delta$ is increased from 0.00 to 0.05 in steps of 0.01. The grey arrow shows how this isochrone would move if the mass to its $1\sigma$ upper limit of 0.207 $M_\odot$.

We see that the [Fe/H] = 0.1 models of mass 0.200 $M_\odot$ are consistent with data at the $1\sigma$ level for $\delta \approx 0.025$, which corresponds to a surface field strength, $B_{surf} \approx 500$ G. If the mass is equal to its $1\sigma$ upper limit, then the models match the data at the $1\sigma$ level for $\delta$ values between 0.018 and 0.033. The corresponding range in surface field strength is 420 – 570 G.

4.3.2 *Analysis based on Kraus et al. measurements*

In figure 8, we show the locations of the two components of AD 3814 in the $T_{eff}$ - $R$ diagram using the results of Kraus et al., together with the non-magnetic 800 Myr isochrones for [Fe/H] = 0.0, 0.1 and 0.2. We see that the primary has a radius that is consistent with the non-magnetic model but is too cool by at least 90 K. The secondary has a temperature that is consistent with the models but is larger by at least $3\sigma$ (i.e. 0.036 $R_\odot$). The black lines show how the models of the components move as $\delta$ is increased. The filled circles correspond from left to right to $\delta$ values of 0.00 to 0.05 in steps of 0.01. We see that increasing [Fe/H] by 0.1 is equivalent to decreasing $\delta$ by ~0.01.

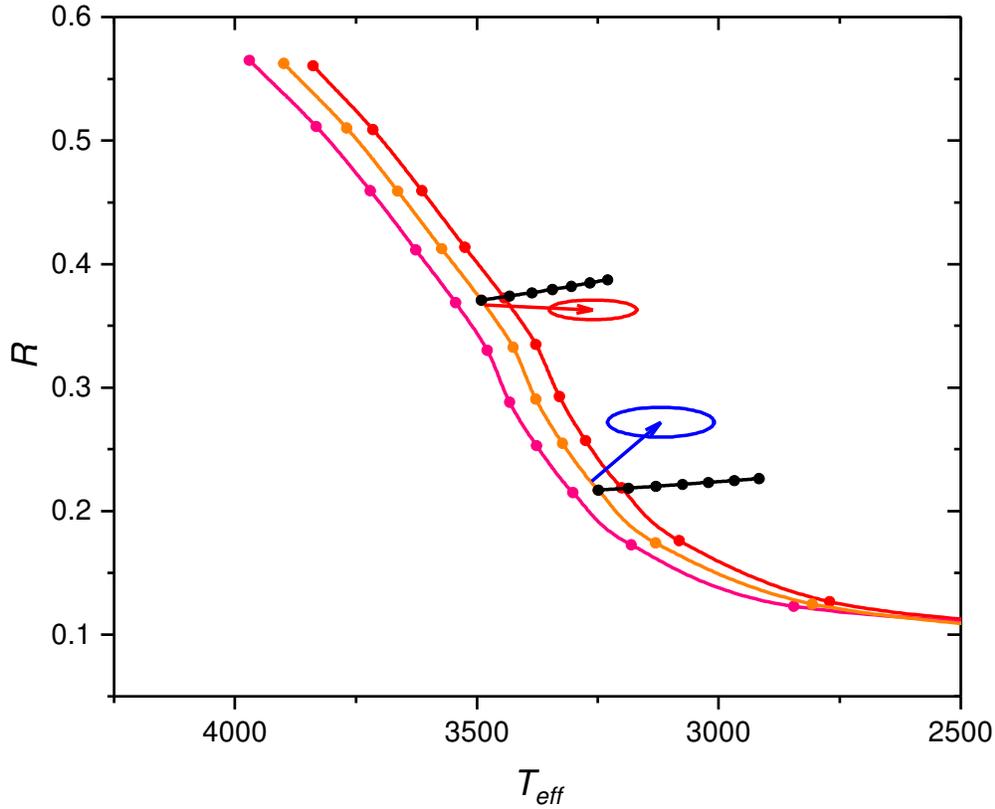

Figure 8. Locations of the primary (red) and secondary (blue) components of AD 3814 in the $T_{eff}$ - $R$ diagram, using the results of Kraus et al. (2017). Also shown are the non-magnetic 800 Myr isochrones for [Fe/H] = 0.0 (pink), 0.1 (orange) and 0.2 (red). The black lines show how the [Fe/H] = 0.1 isochrones moves as $\delta$ is increased for masses 0.20 and 0.40 $M_\odot$. The red and blue arrows show the offsets of the observed mean values of $R$ and $T_{eff}$ from the values predicted by our non-magnetic [Fe/H] = 0.0 models.

  Figure 9 is similar to figure 8 but shows only the location of the primary. The solid and broken ellipses show the 1 and $2\sigma$ errors, respectively, reported by Kraus et al. The black line is the 800 Myr isochrone for magneto-convective models of mass 0.395 $M_\odot$ as $\delta$ is increased from 0.0 to 0.08 in steps of 0.04. We see that the magneto-convective [Fe/H] = 0.1 models of the primary component in AD 3814 match the data at the $2\sigma$ level for $\delta$ in the range 0.01 to 0.05. The corresponding range for the vertical component of the magnetic field strength at the surface of the primary component is 300 to 670 G. Thus, by using the nominal age of Praesepe, we have found that magneto-convective models with interior fields no stronger than 10 kG can replicate the radius and temperature of the primary component of AD 3814, whether we use the results of Kraus et al. (2017) or those of Gillen et al. (2017).

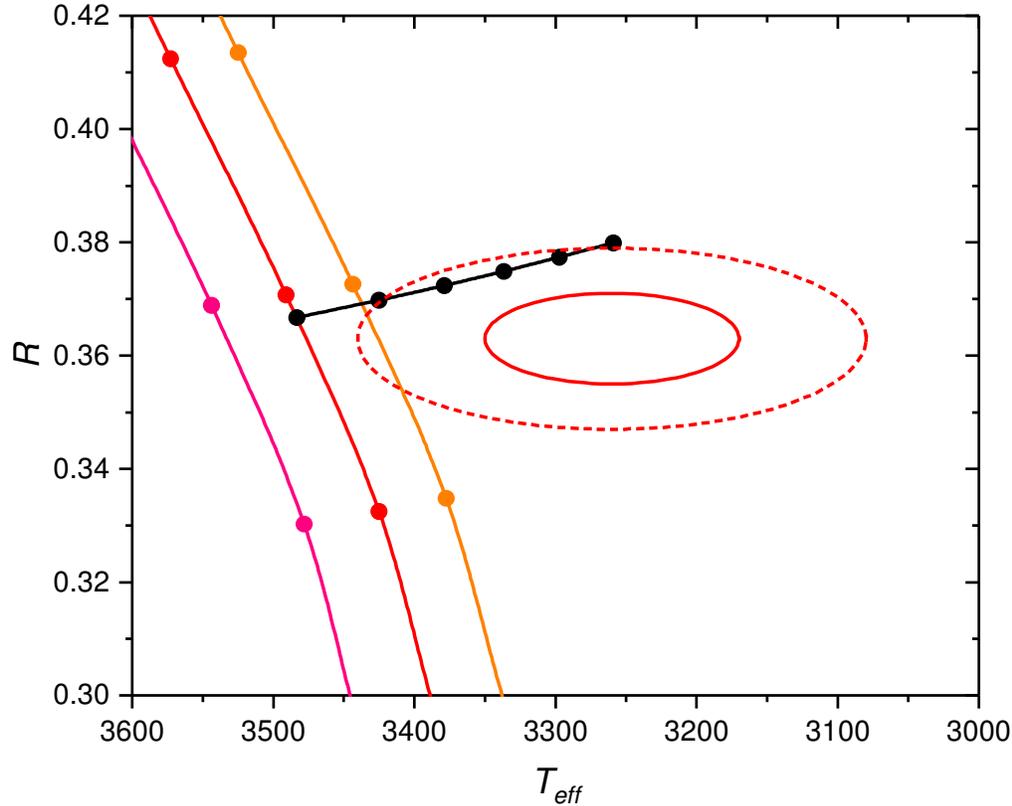

Figure 9. Location of the primary component of AD 3814 in the $T_{eff}$ - $R$ diagram, using the results of Kraus et al. (2017). The solid and broken ellipses show the 1 and $2\sigma$ errors, respectively. Also shown are the non-magnetic 800 Myr isochrones for [Fe/H] = 0.0 (pink), 0.1 (orange) and 0.2 (red). The black line shows how the 0.395 $M_\odot$ [Fe/H] = 0.1 isochrone moves as $\delta$ is increased from 0.0 to 0.05 in steps of 0.01.

In contrast to the results for the primary component, we see from Fig. 6 that our magneto-convective models of the secondary are not consistent with the Kraus et al. data. Kraus et al. have extensively investigated how the presence of dark spots would affect the radius determinations and conclude that the impact on $R_s$ is less than 1.7%, which is insignificant compared to the oversizing of > 13%. However, the presence of dark spots will also impact temperature determinations. The $T_{eff}$ determined by spectroscopy will be weighted towards the hotter regions of the stellar surface and is expected to be larger than $T_{eff}$ determined from radius and bolometric luminosity. Indeed, Kraus et al. find that spectroscopic $T_{eff}$ = 3120 ± 110 K whereas the mean flux from the stellar surface gives $T_{eff}$ = 2970 ± 230 K. If the spots have high contrast or are completely dark, then we can estimate a spot coverage of $f_s$ = 0.18. MM12 have shown that, as regards fitting the empirical stellar properties, there is degeneracy between the effects of spots and the effects of magnetic inhibition of convection. Hence, the isochrones are not changed by the inclusion of spots.

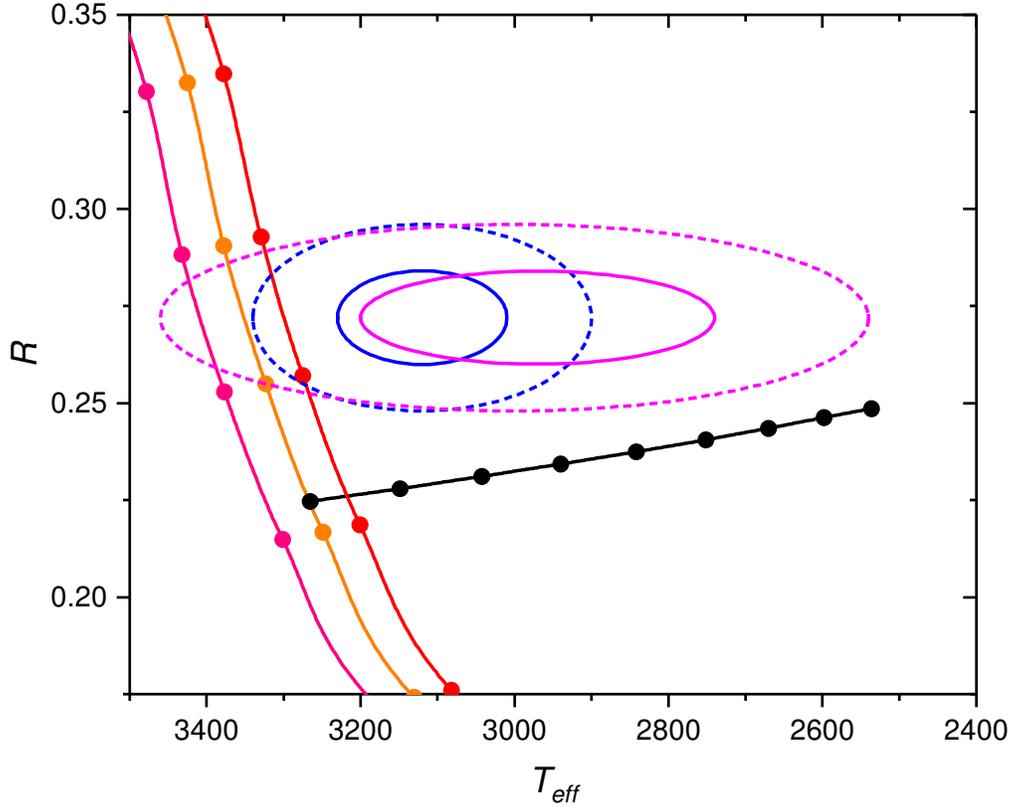

Figure 10. Location of the secondary component of AD 3814 in the $T_{eff}$ - $R$ diagram, using the results of Kraus et al. (2017). The solid and broken ellipses show the 1 and $2\sigma$ errors, respectively, for spectroscopic (blue lines) and bolometric (magenta lines) temperature determinations. Also shown are the non-magnetic 800 Myr isochrones for [Fe/H] = 0.0 (pink), 0.1 (orange) and 0.2 (red). The black line shows how the 0.210 $M_\odot$ [Fe/H] = 0.1 800 Myr isochrone, appropriate to the secondary component of AD 3814, moves as $\delta$ is increased from 0.00 to 0.16 in steps of 0.02.

Figure 10 is similar to figure 9 except the isochrones and error ellipses are for the secondary. The blue and magenta ellipses use the spectroscopic and bolometric $T_{eff}$, respectively. We see that our magneto-convective models at age 800 Myr (which are shown by the black line) do not match the Kraus et al. data, even when the bolometric $T_{eff}$ with $2\sigma$ errors is used.

It has been suggested (Holland et al. 2000; Franciosini et al. 2003) that Praesepe formed from the merger of two clusters of different age and composition. However, Adams et al. (2002) find the observed Praesepe structure is consistent with predictions of numerical simulations, indicating a normal evolutionary history in which no merger has occurred. If there has been a merger, then one or both of the stars in AD 3814 could be younger than 800 Myr. The models show that the secondary component (of mass 0.21 $M_\odot$) stops contracting and reaches the main sequence at age 600 Myr for $\delta = 0$ and 900 Myr for $\delta = 0.08$. Our magneto-convective models for the secondary component are consistent with the measured radius at age 100 Myr for $\delta = 0$ and 200 Myr for $\delta = 0.08$.

In figure 11, we compare the 150 Myr isochrones with the secondary's position in the $T_{eff}$ – $R$ plane. We find that the [Fe/H] = 0.1 models match the data for age 145 ± 25 Myr and $\delta$ = 0.030 ± 0.015. The corresponding field strengths are 550 ± 150 G.

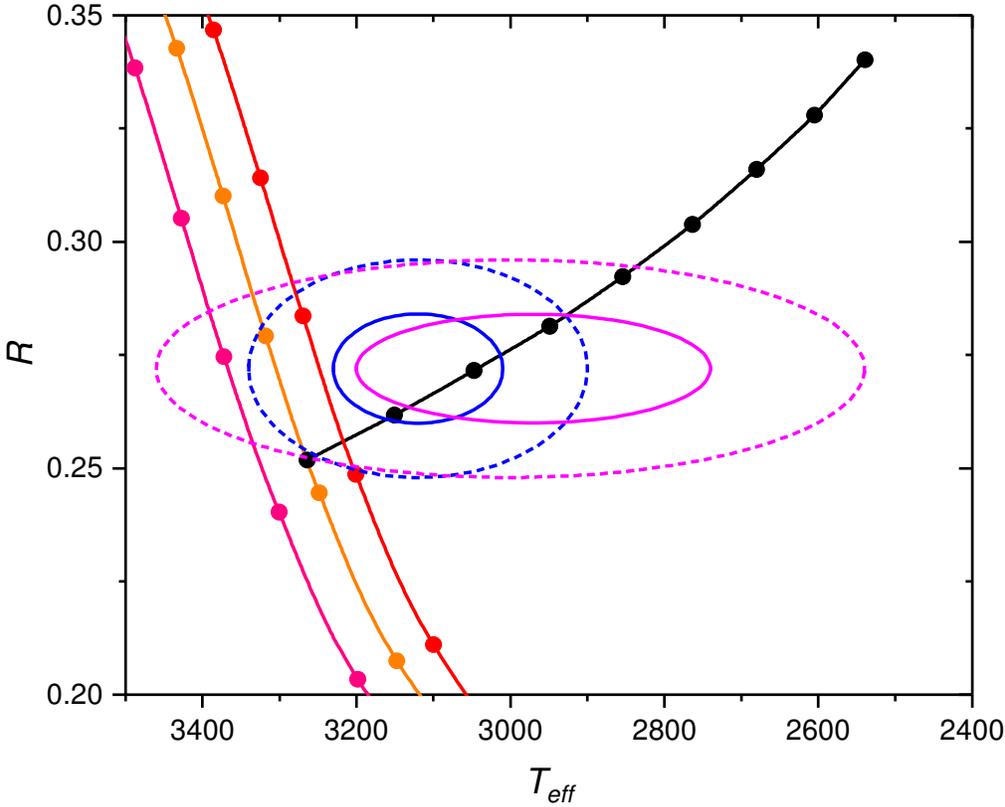

Figure 11. Location of the secondary component of AD 3814 in the $T_{eff}$ - $R$ diagram, using the results of Kraus et al. (2017). The solid and broken ellipses show the 1 and 2$\sigma$ errors, respectively, for spectroscopic (blue lines) and bolometric (magenta lines) temperature determinations. Also shown are the non-magnetic 150 Myr isochrones for [Fe/H] = 0.0 (pink), 0.1 (orange) and 0.2 (red). The black line shows how the 0.210 $M_\odot$ [Fe/H] = 0.1 150 Myr isochrone moves as $\delta$ is increased from 0.00 to 0.16 in steps of 0.02.

If Praesepe was formed by the merger of two (or more) clusters so that some of its stars are significantly younger than 800 Myr, it is natural to ask for other evidence of such stars. One possible candidate is the blue straggler HD 73666, which Fossati et al. (2010) propose is the result of a stellar merger. Abundance analysis by Fossati et al. (2007) indicate that C, N, and O and many other elements are overabundant by about 0.4 dex, which, when compared to the abundance determinations above, supports the suggestion that Praesepe might contain more than one stellar population.

An alternative scenario for a young secondary is that it is itself the result of a stellar merger.

### 4.4 Other systems

In earlier work, we applied our magneto-convective modelling to the following 6 binary systems CM Dra (MM12; MM14), YY Gem (MM14: MacDonald & Mullan 2015), CU Cnc (MM14; MacDonald & Mullan 2015), USco 5 and EPIC 203710387 (MacDonald & Mullan 2017a), and LSPM J1314+1320 (MM17b). In principle, there are 12 stars in these systems, but we deal with only 9 of them: the essentially identical components of YY Gem are treated as a single average component, and only the primary component is treated in CM Dra and LSPM. For all of these systems, we have re-calculated models using the same mixing length ratio, the same outer boundary conditions, and the same magnetic ceiling field of $10^4$ G as used in this paper. With the exception of LP 6613, we have fitted our models to match both the observed $R$ and the observed $T_{eff}$ or $L$ values. For ease of comparison with magnetic field measurements, we have collected our predictions for the surface vertical field strength in Table 1.

Table 1. Predicted magnetic field strengths

| Star | Mass ($M_\odot$) | Radius ($R_\odot$) | [Fe/H] | $\delta$ | $B_{surf}^V$ (G) |
|---|---|---|---|---|---|
| LP 661-13 A | 0.30795 ± 0.00084 | 0.3226 ± 0.0033 | 0.0 | 0.095 ± 0.013 | 1260 ± 80 |
| LP 661-13 B | 0.19400 ± 0.00034 | 0.2174 ± 0.0023 | 0.0 | 0.038 ± 0.015 | 830 ± 250 |
| KELT A | 0.447+0.052-0.047 | 0.540+0.034-0.032 | 0.0 | 0.065 – 0.087[h] | 700 – 810[h] |
|  |  |  |  | 0.023 – 0.053[i] | 410 – 630[i] |
| KELT B | 0.399+0.046-0.042 | 0.453 ± 0.017 | 0.0 | 0.116 – 0.167[h] | 1080 – 1300[h] |
|  |  |  |  | 0.075 – 0.101[i] | 1000 – 1160[i] |
| AD 3814 A | 0.3813 ± 0.0074 | 0.3610 ± 0.0033 | 0.1 | 0.041 – 0.058 | 580 – 720 |
| AD 3814 B | 0.2022 ± 0.0045 | 0.2256+0.0063-0.0049 | 0.1 | 0.018 – 0.033 | 420 – 570 |
| YY Gem[b] | 0.5992 ± 0.0047 | 0.6191 ± 0.0057 | 0.0 | 0.041 – 0.052 | 490 – 550 |
| CM Dra A | 0.23102 ± 0.00089 | 0.2458 ± 0.0019[e] | -0.3[f] | 0.031 – 0.058[j] | 930 – 1270[j] |
|  |  |  | 0.0[g] | 0.000 – 0.021[j] | 0 – 580[j] |
| CU Cnc A | 0.4333 ± 0.0017 | 0.4317 ± 0.0052 | 0.2 | 0.040 – 0.107 | 540 – 880 |
| CU Cnc B | 0.3980 ± 0.0014 | 0.3908 ± 0.0094 | 0.2 | 0.033 – 0.112 | 510 – 940 |
| LSPM A[c] | 0.0885 ± 0.0006 | 0.215 ± 0.018 | 0.0 | 0.013 – 0.051[k] | 440 – 880[k] |
| USco 5 A | 0.3336 ± 0.0022 | 0.862 ± 0.012 | 0.0 | 0.029 – 0.049 | 340 – 440 |
| USco 5 B | 0.3200 ± 0.0022 | 0.852 ± 0.013 | 0.0 | 0.031 – 0.051 | 350 – 450 |
| EPIC A[d] | 0.1183 ± 0.0028 | 0.417 ± 0.010 | 0.0 | 0.026 – 0.055 | 390 – 540 |
| EPIC B[d] | 0.1076 ± 0.0031 | 0.450 ± 0.012 | 0.0 | 0.044 – 0.077 | 480 – 640 |

Notes: [a] KELT = KELT J041621-620046, [b] mean component, [c] LSPM = LSPM J1314+1320, [d] EPIC = EPIC 203710387, [e] Radii of Morales et al. (2009) reduced by 3% due to the presence of polar spots (Morales et al. 2010), [f] abundance and age used by MacDonald & Mullan (2014), [g] abundance and age used by Feiden & Chaboyer (2014), [h] Main Sequence scenario, [i] Pre-Main Sequence scenario, [j] Assumes 17% coverage by completely dark spots, [k] for models with zero magnetic diffusivity.

The two sets of entries in the $\delta$ and $B_{surf}^V$ columns for KELT J041621-620046 correspond to scenarios were the stars are either in the main sequence phase of evolution or the pre-main sequence phase. For CM Dra A, the two sets of entries in the $\delta$ and $B_{surf}^V$ columns are from models using different age and [Fe/H] estimates. The upper entries are based on our models that use an age of

4.1 ± 0.8 Gyr for the white dwarf proper motion companion (Morales et al. 2009) and [Fe/H] = −0.30 ± 0.12 (Terrien et al. 2012). Feiden & Chaboyer (2014) have argued that the age of the white dwarf companion had been underestimated and a more realistic age is 8.5 ± 3.5 Gyr. They also propose that CM Dra has a near-solar [Fe/H] based on the possibility that it is a member of the thick disk, and therefore likely enriched in $\alpha$-elements. The lower entries for CM Dra A in the table correspond to the models that we obtain when we use the age and [Fe/H] values proposed by Feiden & Chaboyer (2014).

## 5. DISCUSSION AND CONCLUSION

We have obtained magneto-convective models of 15 low-mass stars in which magnetic fields are strong enough to have a detectable effect on the structural properties of each star. In particular, the most prominent empirical effect that we seek to replicate by our models is a statistically significant increase in the radius of the star compared to standard (non-magnetic) models. The 15 stars which are the subject of the present paper have masses in the range 0.09-0.6 $M_\odot$, i.e. they span the range where main sequence stars are believed to undergo a transition to complete convection.

In obtaining the new magneto-convective models reported here, the principal difference from models we have reported previously for low-mass stars is the following: We have capped the magnetic field strength in the interior of each star at a value of 10 kG. This is a much weaker field, by factors of $10^{3-4}$, than was reported in our earliest paper (MM01): we now recognize that the strong interior fields reported in MM01 are untenable. Moreover, those strong fields are now seen to be unnecessary: they were obtained in an ill-informed attempt to interpret coronal data for low-mass stars by proposing that complete convection does *not* set in throughout a star with a mass smaller than (roughly) 0.3 $M_\odot$. The onset of complete convection in such stars, proposed theoretically by Limber (1958), has recently received empirical support from *chromospheric* data (Houdebine et al. 2017). In the present paper, our reason for setting a cap of on field strength is that fields in the range 10-20 kG have been predicted to exist inside low-mass stars based on 3D numerical models of turbulent dynamos. Despite the apparent smallness of the 10 kG limit, the magneto-convective models presented here indicate that such fields nevertheless lead to measurable structural changes in low-mass stars.

In our 15 target stars, we have found in 14 cases magneto-convective solutions that replicate the radius measurements when we use evolutionary ages that are consistent with independent age estimates.

The 15th star is the secondary component of AD 3814 = PTFEB132.707+19.810 (a member of the Praesepe cluster). For this object, Gillen et al. (2017) and Kraus et al. (2017) find significantly discrepant results for the radius, with Kraus et al. finding a radius 17% larger than that found by Gillen et al. Both groups measure similar effective temperatures that are at least 130 K cooler than non-magnetic models. Based on the results of Gillen et al. (2017), we find that, using the canonical age of Praesepe, the temperature and radius of AD3814 B can be matched by magnetic models with $\delta \approx 0.025$, which corresponds to a surface field strength,

$B_{surf} \approx 500$ G. However, our magnetic models do not match the radius found by Kraus et al. if we confine ourselves to the canonical age of Praesepe. Instead, an age of about 150 Myr is required, which is several times smaller than the canonical age of Praesepe.

The vertical magnetic field strengths at the *surfaces* of all 15 stars for which we have obtained magneto-convective models are found to be in the range from 300 to 1300 G. The total field strength on the surface of any star will exceed these values by amounts that depend on the actual field geometry: if the fields happen to be isotropic, then the total field strengths would be of order √3 times stronger, i.e. 500 to 2300 G. But if (as reported by Shulyak et al. 2014) the magnetic field is dominated by its radial (i.e. vertical) component, then the total field strength will be comparable to 300 - 1300 G. Whichever of these two limits are closer to reality, the numerical values are consistent with the empirical ranges which have been reported for fields in active low-mass stars (0.1-7 kG [Shulyak et al. 2017]). In principle, our predictions can at some point in time be tested, e.g. by the ongoing BinaMics project (Neiner et al. 2015) which includes a number of low mass binary stars, including YY Gem, in its target list.

Finally, we note that inclusion of dark spots or a higher magnetic field ceiling would result in lower predicted values for the surface magnetic field.

Acknowledgements

Partial support for this work is provided by the Delaware Space Grant.

REFERENCES

Adams, J. D. et al. 2002, AJ, 124, 1570
Bass, G., Orosz, J. A., Welsh, W. F., et al. 2012, ApJ, 761, 157
Boesgaard, A. M., Roper, B. W., & Lum, M. G. 2013, ApJ, 775, 58
Brandt, T. D., & Huang, C. X. 2015, ApJ, 807, 24
Browning, M. K. 2008, ApJ, 676, 1262
Browning, M. K., Weber, M. A., Chabrier, G., & Massey, A. P. 2016, ApJ, 818, 189
Carrera, R., & Pancino, E. 2011, A&A, 535, A30
Carter J. A. et al., 2011, Sci, 331, 562
Cattaneo, F. & Hughes, D. W. 1996, Phys Rev E, 54, R4532
Chabrier, G., Gallardo, J., & Baraffe, I. 2007, A&A, 472, L17
Charbonneau, P. & MacGregor, K.B. 1997, ApJ, 486, 502
Choudhuri, A. R. & Gilman, P. A. 1987, ApJ 316, 788
Delorme, P., Collier Cameron, A., Hebb, L., et al. 2011, MNRAS, 413, 2218
Dittmann, J. A., Irwin, J. M., Charbonneau, D., et al. 2017, ApJ, 836, 124
Dupuy, T. J., Forbrich, J., Rizzuto, A., et al. 2016, ApJ, 827, 23
Durney, B. R., De Young, D. S., & Roxburgh, I. W. 1993, Solar Phys., 145, 207
Feiden, G. A. & Chaboyer, B. 2012, ApJ, 761, 30 (FC)

Feiden, G. A. & Chaboyer, B. 2013, ApJ, 779, 183 (FC13)
Feiden, G. A., & Chaboyer, B. 2014, A&A, 571, A70
Feiden, G. A., Chaboyer, B., & Dotter, A., 2011, ApJ, 740, L25
Fleming, T. A., Giampapa, M. S., Schmitt, J., & Bookbinder, J. A. 1993, ApJ, 410, 387
Fossati, L., Bagnulo, S., Monier, R., et al. 2007, A&A, 476, 911
Fossati, L., Mochnacki, S., Landstreet, J., & Weiss, W. 2010, A&A, 510, A8
Franciosini, E., Randich, S., & Pallavicini, R. 2003, A&A, 405, 551
Friel, E. D., & Boesgaard, A. M. 1992, ApJ, 387, 170
Gillen, E., Hillenbrand, L. A., David, T. J., et al. 2017, ArXiv e-prints, arXiv:1706.03084
Gough, D. O., & Tayler, R. J. 1966, MNRAS, 133, 85
Han, E. et al. 2017, AJ, 154, 100
Holland, K., Jameson, R. F., Hodgkin, S., Davies, M. B., & Pinfield, D. 2000, MNRAS, 319, 956
Houdebine, E. R., & Mullan, D. J. 2015, ApJ, 801, 106
Houdebine, E. R., Mullan, D. J., Bercu, B., Paletou, F., & Gebran, M. 2017, ApJ, 837, 96
Jackson, J. D. 1962, Classical Electrodynamics, pp. 141-145, pp. 156-158.
Kharchenko, N. V., Piskunov, A. E., Roeser, S., Schilbach, E., & Scholz, R.-D. 2005, A&A, 438, 1163
Kochukhov, O., & Lavail, A. 2017, ApJL, 835, L4
Kraus, A. L., Tucker, R. A., Thompson, M. I., Craine, E. R., & Hillenbrand, L. A. 2011, ApJ, 728, 48
Kraus, A. L., Douglas, S. T., Mann, A. W., et al. 2017, ApJ 845, 72
Krishna-Swamy, K. S. 1966, ApJ, 145, 174
Leggett, S. K., Allard, F., Dahn, C., et al. 2000, ApJ, 535, 965
Lépine, S., Hilton, E. J., Mann, A. W., et al. 2013, AJ, 145, 102
Limber, D. N. 1958, ApJ, 127, 387
Lubin, J. B., Rodriguez, J. E., Zhou, G., et al. 2017, ApJ 844, 134
MacDonald, J. 2015, The Structure and Evolution of Single Stars: An Introduction, IOP Concise Physics, Morgan & Claypool
MacDonald, J., & Mullan, D. J. 2012, MNRAS, 421, 3084 (MM12)
MacDonald, J., & Mullan, D. J. 2013, ApJ, 765, 126
MacDonald, J., & Mullan, D. J. 2014, ApJ, 787, 70 (MM14)
MacDonald, J., & Mullan, D. J. 2017a, ApJ, 834, 67 (MM17a)
MacDonald, J., & Mullan, D. J. 2017b, ApJ, 843, 142 (MM17b)
Mann, A. W., Feiden, G. A., Gaidos, E., Boyajian, T., & von Braun, K. 2015, ApJ, 804, 64
McLean, M., Berger, E. Irwin, J., et al. 2011 ApJ, 741, 27
Morales, J. C., Ribas, I., Jordi, C., et al. 2009, ApJ, 691, 1400
Morales, J., Gallardo, J., Ribas, I., Jordi, C., Baraffe, I., & Chabrier, G. 2010, ApJ, 718, **502**
Morin, J., Donati, J.-F., Petit, P. et al 2010, MNRAS 407, 2269
Mullan, D. J. 2009, Physics of the Sun: a First Course, Boca Raton: CRC Press, p. 86
Mullan, D. J., & MacDonald, J. 2001, ApJ, 559, 353 (MM01)

Mullan, D. J., MacDonald, J., & Rabello-Soares, M. C. 2012, ApJ, 755, 79
Neiner, C., Morin, J., & Alecian, E. 2015, SF2A-2015: Proceedings of the Annual meeting of the French Society of Astronomy and Astrophysics, Eds. F. Martins, S. Boissier, V. Buat, L. Cambrésy, P. Petit, 213
Pace, G., Pasquini, L., & François, P. 2008, A&A, 489, 403
Reid, I. N., Hawley, S. L., & Gizis, J. E. 1995, AJ, 110, 1838
Ribas, I. 2006, Ap&SS 304, 89
Roberts, P. H., & King, E. M. 2013, Rep. Progr. Phys. 76, 096801
Saar, S. H. 1996, IAU Symp. No. 176, eds. K. G. Strassmeier and J. L. Linsky, Kluwer Academic Publishers, Dordrecht, p. 237
Schaeffer, N., Jault, D., Nataf, H. C., & Fournier, A. 2017, Geophys. J. Intern., 211, 1
Spada, F., Demarque, P., Kim, Y.-C., & Sills, A. 2013, ApJ, 776, 87
Spruit, H. C. 1992, in Surface Inhomogeneities on Late-type Stars, ed. P. B. Byrne & D. J. Mullan (Lecture Notes in Physics, Vol. 397; Berlin: Springer), 78
Shulyak, D., Reiners, A., Seemann, U., et al. 2014, A&A, 563, A35
Shulyak, D., Reiners, A., Engeln, A., et al. 2017, Nature Astron. Lett., 1, 0184
Taylor, B. J. 2006, AJ, 132, 2453
Terrien, R. C., Fleming, S. W., Mahadevan, S., et al. 2012, ApJL, 760, L9
Torres, G., & Ribas, I. 2002, ApJ, 567, 1140
Yadav, R. K., Christensen, U. R., Morin, J., et al. 2015, ApJ, 813, L31
Yang, X. L., Chen, Y. Q., & Zhao, G. 2015, AJ, 150, 158
Zahn J.-P., 1989, A&A, 220, 112
Zhang, K., Chan, K. H., Zou J. et al. 2003, ApJ, 596, 663
Zhang, K., & Schubert, G. 2006, Rep. Progr. Phys., 69, 1581